\documentclass[aps,11pt,nofootinbib,preprintnumbers,showpacs]{revtex4}

\usepackage{amssymb, amsfonts, amsmath, bm}
\usepackage{amsfonts}
\usepackage{ulem}
\usepackage[dvipdfmx]{graphicx,color}

\begin{document}




\title{
A Kaluza-Klein black lens in five dimensions }
\vspace{2cm}

\author{Shinya Tomizawa\footnote{tomizawasny@stf.teu.ac.jp}
} 
\vspace{2cm}
\affiliation{
 Department of Liberal Arts, Tokyo University of Technology, 5-23-22, Nishikamata, Otaku, Tokyo, 144-8535, Japan
}




\begin{abstract} 
We obtain a supersymmetric Kaluza-Klein black lens solution in Taub-NUT space in the five-dimensional minimal ungauged supergravity. It is shown that the spacetime has a degenerate horizon with the spatial cross section of the lens space topology $L(n,1)=S^3/\mathbb{Z}_n$ and looks like the four-dimensional Minkowski spacetime in the neighborhood of spatial infinity.  In contrast to the horizon topology, from a five-dimensional point of view, the spatial infinity has the topology of $S^3$ rather than the lens space, for which this solution has an asymptotically flat limit.  We discuss several properties of such a black lens, in particular,  the effect by the compactification of an extra-dimension and some physical differences from the asymptotically flat supersymmetric black lens which has recently been found. 
\end{abstract}
\date{\today}
\maketitle




\section{Introduction}\label{sec:1}

Higher dimensional black holes/rings and other extended black objects have been considered to play essential roles in the various context of the statistical counting of black-hole entropy, the AdS/CFT correspondence, and the black-hole production at an accelerator. 
In particular, physics of black holes in five-dimensional Einstein-Maxwell-Chern-Simons (EMCS) theory has recently been the subject of increased attention, since the five-dimensional EMCS theory describes the bosonic sector of five-dimensional minimal supergravity as a low-energy limit of string theory, as well as one of the simplest theories of supersymmetry. 
So far,  several types of black hole solutions in this theory have been found by using recent development of solution generating techniques~\cite{Gauntlett:2002nw,Mizoguchi:2011zj,Mizoguchi:2012vg,Tomizawa:2012nk,Ford:2007th,Giusto:2007tt,Compere:2009zh,Bouchareb:2007ax,Galtsov:2008jjb} and they have been classified in the context of the uniqueness theorems~\cite{Tomizawa:2009ua,Tomizawa:2009tb,Tomizawa:2010xj}. 
However, it is evident that the construction of all black hole solutions has not been achieved yet. 

\medskip
It is now evident that even in vacuum Einstein theory, there is a much richer variety of black hole solutions in higher dimensions.
For instance, the topology theorem~\cite{Galloway:2005mf,Cai:2001su,Hollands:2007aj,Hollands:2010qy} for stationary black holes which is generalized to five dimensions allows the topology of the spatial cross section of the event horizon to be either a sphere $S^3$, a ring $S^1\times S^2$ or lens spaces $L(p,q)$ under the assumptions of asymptotic flatness and bi-axisymmetry. 
As for both of the sphere $S^3$ and the ring topology $S^1\times S^2$, the corresponding exact solutions have been found as stationary solutions to the five-dimensional vacuum Einstein  equations~\cite{Tangherlini:1963bw,Myers:1986un,Emparan:2001wn,Pomeransky:2006bd}. 
However, for the lens space topologies $L(p,q)$, they have not yet been found in spite of efforts for some researchers to construct them as a regular vacuum solution~\cite{Evslin:2008gx,Chen:2008fa}.

\medskip
Recently, however, within the class of the supersymmetric solutions in the bosonic sector of the five-dimensional minimal ungauged supergravity,  the first regular black lens solution with asymptotic flatness has been constructed  by Kunduri and Lucietti~\cite{Kunduri:2014kja} for the horizon topology of $L(2,1)=S^3/{\mathbb Z}_2$. Thereafter, this has been generalized to the supersymmetric solution with the horizon of the more general lens space topologies $L(p,1)=S^3/{\mathbb Z}_p\ (p\ge 3)$ by the author of this paper and Nozawa~\cite{Tomizawa:2016kjh}.   
How to construct these solutions is based on the well-known work of the classification of supersymmetric solutions by Gauntlett {\it et. al}~\cite{Gauntlett:2002nw}. 
Moreover, this has been immediately extended to a multi-black lens solution~\cite{Tomizawa:2017suc} and a cosmological black lens solution~\cite{Tomizawa:2017uxp}.

\medskip
The assumption of asymptotic flatness is mainly related to the context of a braneworld model,  where the size of higher dimensional black holes can become much smaller than the size of extra-dimensions.
However, since our visible world is thought to be macroscopically four-dimensional, extra-dimensions must be compactified in some sense. 
In this direction, it has been considered to be of great interest to consider higher dimensional Kaluza-Klein black holes since they look like four-dimensional, at least, at infinity although they appear higher dimensional near the horizon~\cite{Tomizawa:2011mc,Dobiasch:1981vh,Rasheed:1995zv,Ishihara:2005dp,Ishihara:2006iv,Nakagawa:2008rm,Tomizawa:2008hw,Matsuno:2008fn,Tomizawa:2008rh}. 
Such Kaluza-Klein solutions can be expected to help us to get some insights into the major open problem about how to compactify and stabilize extra-dimensions in string theory. The main purpose of this paper is to construct a certain type of supersymmetric Kaluza-Klein black lens solutions in five-dimensional minimal supergravity,  understanding the novel effect by the compactification of an extra-dimension and making it clear what is essentially different from the asymptotic flat black lens solutions.

\medskip
It is mathematically  well-known that the lens spaces $L(p,q)$ (, where $p$ and $q$ are coprime integers) are quotients of $S^3$ by ${\mathbb Z}/p$-action, which can be regarded as an $S^1$ fiber bundle over an $S^2$.  In particular, the regular metric on the lens space $L(p,1)=S^3/{\mathbb Z}_p$ is simply written as
\begin{eqnarray}
ds^2=\frac{1}{4}\left[R_1^2\left(\frac{d\psi}{p}+\cos\theta d\phi\right)^2+R_2^2(d\theta^2+\sin^2\theta d\phi^2)\right], \label{eq:lens}
\end{eqnarray}
where $0\le\psi<4\pi$, $0\le\phi<2\pi$, $0\le \theta\le \pi$, and $R_1/2$, $R_2/2$ are the radii of the $S^1$ and $S^2$, respectively. 
When $p=1$, this reduces to the metric on an $S^3$ written in terms of the Euler angles, which is often refereed as round $S^3$ for $R_1=R_2$ and  squashed $S^3$ for $R_1\not= R_2$, respectively. 
An asymptotically flat black lens spacetime~\cite{Kunduri:2014kja,Tomizawa:2016kjh} has the spatial infinity of a round $S^3$, where the ratio $R_2/R_1$ is $1$.  
In this paper, we would like to consider an asymptotically Kaluza-Klein black lens spacetime with the spatial infinity of a squashed $S^3$ ($p=1$ in Eq.~(\ref{eq:lens})) and a horizon of the lens space $L(n,1)$ ($p=n$ in Eq.~(\ref{eq:lens})), where the size of an $S^1$ is much small than that of an $S^2$ (i.e., $R_1/R_2 \to0$) at infinity and $R_1$ and $ R_2$ are finite on the horizon.  
Therefore, we will impose the appropriate boundary conditions on the parameters included in the supersymmetric solutions on the Taub-NUT space.

\medskip
This paper is organized as follows. 
In the following section~\ref{sec:solution}, we give the supersymmetric black lens solutions with bubbles in Taub-NUT space in the five-dimensional minimal ungauged supergravity. 
In section~\ref{sec:boundary}, we impose the boundary conditions so that the spacetime is asymptotically Kaluza-Klein spacetime,  has no closed timelike curves  (CTCs) appear around the horizon, no conical and curvature singularities in the domain of outer communications, and no orbifold singularities nor Dirac-Misner strings on the axis.  
In Section~\ref{sec:analysis}, we discuss  some physical properties of such a black lens.     
In the final section~\ref{sec:discuss}, we devote ourselves to the summary and discussion on our results.




\section{solution}
\label{sec:solution}

We consider supersymmetric solutions in the five-dimensional minimal ungauged supergravity, whose bosonic sector is described  by the Lagrangian of the Einstein-Maxwell- Chern-Simons theory with a special coupling constant~\cite{Gauntlett:2002nw}
\begin{eqnarray}
\mathcal L=R \star 1 -2 F \wedge \star F -\frac 8{3\sqrt 3}A \wedge F \wedge F \,, 
\end{eqnarray}
where $F=d A$ is the Maxwell field. 
 The metric and the gauge potential $1$-form of a supersymmetric Kaluza-Klein black lens solution in this theory have a following form 
\begin{eqnarray}
\label{metric}
ds^2&=&-f^2(dt+\omega)^2+f^{-1}ds_{M}^2,\\
A&=&\frac{\sqrt 3}{2} \left[f(d t+\omega)-\frac KH(d \psi+\chi)-\xi \right]\,,
\end{eqnarray}
where $ds^2_M$ is chosen to be the metric on the Gibbons-Hawking space (, which is more precisely called multi-Taub-NUT space)~\cite{Gibbons:1979zt},
\begin{eqnarray}
ds^2_M&=&H^{-1}(d\psi+\chi)^2+H[dr^2+r^2(d\theta^2+\sin^2\theta d\phi^2)], \\
H&=&h_0+\sum_{i=1}^n\frac{h_i}{r_i}:=h_0+\frac{n}{r_1}-\sum_{i=2}^n\frac{1}{r_i}, \\ \label{Hdef}
d \chi &=& * d H \,, \label{eq:chi0}
\end{eqnarray}
where $r_i:=|{\bm r}-{\bm r_i}|=\sqrt{r^2-2z_ir\cos\theta+z_i^2}$ with $z_i\ (i=1,\cdots,n)$ are constants.
It should be noted that the harmonic functions of this solution differ from that of the asymptotically flat black lens~\cite{Kunduri:2014kja,Tomizawa:2016kjh} by the presence of the constant $h_0$ in $H$, which changes the asymptotic structure from asymptotic flatness into asymptotic Kaluza-Klein.

Furthermore, the other elements are given by~\cite{Gauntlett:2002nw} 
\begin{eqnarray}
f^{-1}&=&H^{-1}K^2+L,\\
\omega&=&\omega_\psi(d\psi+\chi)+\hat \omega,\\
\omega_\psi&=&H^{-2}K^3+\frac{3}{2} H^{-1}KL+M, \\
*d\hat\omega&=&HdM-MdH+\frac{3}{2}(KdL-LdK),\label{eq:hatomega0}\\
d \xi& =&-* d K \,,\label{eq:xi0}
\end{eqnarray}
where the functions $K,L$ and $M$ are harmonic functions on three-dimensional Euclid space, which are given by
\begin{eqnarray}
M&=&m_0+\sum_{i=1}^n\frac{m_i}{r_i},\label{Mdef}\\
K&=&k_0+\sum_{i=1}^n\frac{k_i}{r_i},\label{Kdef}\\
L&=&l_0+\sum_{i=1}^n\frac{l_i}{r_i}.\label{Ldef}
\end{eqnarray}
From Eqs.~(\ref{eq:chi0}) and (\ref{eq:hatomega0}) and  (\ref{eq:xi0}), the 1-forms ($\chi, \xi, \hat \omega$) are determined to give
\begin{eqnarray}
\chi&=&\sum_{i=1}^nh_i\tilde\omega_i,\\
\xi&=-&\sum_{i=1}^nk_i\tilde\omega_i,\\
\hat \omega&=&\sum_{i,j=1(i\not=j)}^n\left(h_im_j+\frac{3}{2}k_i l_j \right)\hat \omega_{ij}-\sum_{i=1}^n\left(m_0h_i+\frac{3}{2}l_0k_i-h_0m_i\right)\tilde\omega_i+cd\phi,
\end{eqnarray}
where the 1-forms $\tilde\omega_i$ and $\hat\omega_{ij}$ are 
\begin{eqnarray}
\tilde\omega_i&=&\frac{r\cos\theta-z_i}{r_i}d\phi,\\
\hat\omega_{ij}&=&\frac{r^2-(z_i+z_j)r\cos\theta+z_iz_j}{z_{ji}r_ir_j}d\phi,
\end{eqnarray}
where $z_{ji}:=z_j-z_i$ and $c$ is a constant. 
The  Killing vector $\partial/\partial \psi$ in the Gibbons-Hawking space  is also a symmetry generator for the five-dimensional metric $g_{\mu \nu}$ and the gauge field $A_\mu$.

It should be noted that there exists a gauge freedom of redefining harmonic functions~\cite{Bena:2005ni} 
\begin{eqnarray}
K\to K+aH,\qquad L\to L-2aK-a^2H,\qquad M\to M-\frac{3}{2}aL+\frac{3}{2}a^2K+\frac{1}{2}a^3H,\label{eq:trans}
\end{eqnarray}
where $a$ is a constant. Under the transformation (\ref{eq:trans}),  
the constant term $k_0$ in $K$ changes as 
$k_0\to k_0+ ah_0$.  
In an appropriate choice of $a$,  one can put
\begin{eqnarray}
k_0=0.  \label{eq:k0}
\end{eqnarray}
In the limit of $h_0\to0$, this solution reduces to the asymptotically flat black lenses~\cite{Tomizawa:2016kjh} (in the appropriate choice of the parameters),  the BMPV black hole~\cite{Breckenridge:1996is} with the horizon of spherical topology for $n=1$, and  the supersymmetric black lens with the horizon of the lens space topology $L(2,1)=S^3/{\mathbb Z}_2$ of Kunduri-Lucietti~\cite{Kunduri:2014kja} for $n=2$.




\section{Boundary conditions}\label{sec:boundary}

In order to obtain a supersymmetric Kaluza-Klein black lens solution, we impose  boundary conditions at spacetime boundaries,  at infinity $r\to\infty$,  on the horizon ${\bm r}={\bm r}_1$,  at the points ${\bm r}={\bm r}_i\ (i=2,...,n)$, and on the $z$-axis $x=y=0$ in ${\mathbb E}^3$ of the Gibbons-Hawking base space.  
\begin{description}
\item(i)  At infinity $r\to\infty$, the extra-dimension of the spacetime is compactified so that the size of the fifth dimension is much smaller than that of the other spatial dimensions.  Hence,  the spacetime can be asymptotically approximated as an $S^1$ fiber bundle over four-dimensional Minkowski spacetime. 
Moreover, we make an additional assumption that spatial infinity is topologically $S^3$ so that the obtained solution can have a limit to an asymptotically flat solution.
\item(ii) at the horizon ${\bm r}={\bm r}_1$, the surface should be a smooth degenerate null surface whose spatial cross section has a topology of the lens space 
$L(n,1)=S^3/{\mathbb Z}_n$.  
\item(iii) at the $(n-1)$ points ${\bm r}={\bm r}_i\ (i=2,...,n)$, where each harmonic function diverges, the metric behaves like the origin of the Minkowski spacetime. 
\item(iv)  On the $z$-axis in ${\mathbb E}^3$ of the Gibbons-Hawking base space, there exist no Dirac-Misner strings, and orbifold singularities. 
\end{description}
Furthermore,  under these boundary conditions,  the spacetime is required to allow neither CTCs nor (conical and curvature) singularities.




\subsection{Infinity}
\label{sec:bc_infinity}

For $r \to \infty$, the metric functions $f$ and  $\omega_\psi$ behave, respectively, as 
\begin{eqnarray}
f^{-1}&\simeq &l_0+\frac{\sum_il_i}{r}, \qquad 
\omega_\psi\simeq m_0+\frac{\frac{3}{2}l_0\sum_{i} k_i+\sum_im_i}{r}. 
\end{eqnarray}
The $1$-forms $\omega$ and $\chi$ behave, respectively, as 
\begin{eqnarray}
\chi&\simeq&\cos\theta d\phi
\\
\omega
                    &\simeq &m_0\left(d\psi +\sum_ih_i\cos\theta d\phi \right)-\sum_{i}\left(m_0h_i+\frac{3}{2}l_0k_i-h_0m_i\right)\cos\theta d\phi \notag \\
                    &&+\left(\sum_{i,j(i\not =j)}\frac{h_im_j
                    +\frac{3}{2}k_il_j}{z_{ji}} +c\right)d\phi.
\end{eqnarray}
Therefore, the metric can be approximated as
{\small
\begin{eqnarray}
ds^2&\simeq& -l_0^{-2}\Biggl[dt+m_0\left(d\psi +\cos\theta d\phi \right)\notag\\
                      &&-\sum_{i}\left(m_0h_i+\frac{3}{2}l_0k_i-h_0m_i\right)\cos\theta d\phi 
                    +\left(\sum_{i,j(i\not =j)}\frac{h_im_j+\frac{3}{2}k_il_j}{z_{ji}} +c\right)d\phi\Biggr]^2\notag\\
                  &&  +l_0\left[h_0^{-1} \left(d\psi+\sum_ih_i \cos\theta d\phi \right)^2+h_0\left\{dr^2+r^2(d\theta^2+\sin^2\theta d\phi^2)\right\} \right].
\end{eqnarray}
}
The boundary condition (i) at infinity demands the parameters to satisfy
\begin{eqnarray}
&&l_0=1,\label{eq:l0}\\
&&\sum_{i}\left(m_0h_i+\frac{3}{2}l_0k_i-h_0m_i\right)=0,\label{eq:m0}\\
&&c=-\sum_{i,j(i\not =j)}\frac{h_im_j+\frac{3}{2}k_il_j}{z_{ji}}.\label{eq:c}
\end{eqnarray}
In the choice of these parameters,  for $r\to \infty$, the metric asymptotically becomes 
\begin{eqnarray}
ds^2&\simeq&-\left[dt+m_0(d\psi+\cos\theta d\phi)\right]^2+h_0\left[d r^2+  r^2(d\theta^2+\sin^2\theta d\phi^2)\right] +h_0^{-\frac{1}{2}}\left(\psi+\cos\theta d\phi\right)^2\\
&=& -\left[dt+m_0(d\psi+\cos\theta d\phi)\right]^2+d\bar r^2+ \bar r^2(d\theta^2+\sin^2\theta d\phi^2) +\left(dw+h_0^{-\frac{1}{2}}\cos\theta d\phi\right)^2,
\end{eqnarray}
where we have defined $\bar r=h_0^{\frac{1}{2}}\ r$ and $w=h_0^{-\frac{1}{2}}\ \psi$. 
Because of the presence of the cross term $-m_0(d\psi+\cos\theta d\phi)dt$, this does not seem to satisfy the boundary condition (i) but this is because the asymptotic form is not at rest frame. 
Therefore, to move to the rest frame, let us define the coordinates $(\bar t, \bar w)$ by
\begin{eqnarray}
t=\sqrt{1-h_0m_0^2}\ \bar t,\quad w=\frac{\bar w+h_0^{-\frac{1}{2}}m_0 \bar t}{\sqrt{1-h_0m_0^2}}. 
\end{eqnarray}
In terms of these new coordinates, the asymptotic metric can be rewritten as
\begin{eqnarray}
 ds^2&\simeq& -dt^2+d\bar r^2+ \bar r^2(d\theta^2+\sin^2\theta d\phi)^2+\left(d\bar w+h_0^{-\frac{1}{2}}\cos\theta d\phi\right)^2.
\end{eqnarray}
This is the metric of an $S^1$ fiber bundle over Minkowski spacetime, where the radius of $S^2$ becomes infinite but that of $S^1$ remains finite. 
The absence of conical singularities requires the range of angles to be $0\le \theta \le \pi$, $0\le \phi <2\pi $ and $0\le \psi <4\pi$ with the identification $\phi\sim \phi+2\pi$ and $\psi\sim \psi+4\pi$, under which assumptions the spatial infinity  is topologically an $S^3$ rather than a lens space. 
From $w\sim w +2\pi R_k$, $\bar w\sim \bar w +2\pi R_5$, we can observe that radius of the Kaluza-Klein circle $R_5$ at infinity is
\begin{eqnarray}
R_5=\sqrt{1-h_0m_0^2} R_k=\frac{2\sqrt{1-h_0m_0^2}}{\sqrt{h_0}}.
\end{eqnarray}
In the limit $h_0\to 0$, the Kaluza-Klein circle $R_5$, namely, the size of an extra-dimension becomes infinite, which corresponds to the asymptotically flat black lens~\cite{Kunduri:2014kja,Tomizawa:2016kjh}.




\subsection{Horizon ${\bm r}={\bm r}_1$}
First, we see that the point source ${\bm r}={\bm r}_1$ in the harmonic functions $H,K,L$, and $M$ corresponds to a degenerate Killing horizon whose topology of the spatial cross section is the lens space of $L(n,1)=S^3/\mathbb{Z}_n$. 
Since without loss of generality, we can choose the point source ${\bm r}_1$ as the origin of ${\mathbb E}^3$ in the Gibbons-Hawking base space, we here consider only the geometry near the origin $r=0$.  
Near this point, the functions $f^{-1}$ and $\omega_\psi$ are expanded as
\begin{eqnarray}
f^{-1}\simeq \frac{k_1^2/h_1+l_1 }{r}+c_1',\qquad \omega_\psi &\simeq&\frac{k_1^3/h_1^2+3k_1l_1/2n+m_1}{r}+c_2',
\end{eqnarray}
where we have defined the constants $c_1'$ and $c_2'$ by
\begin{eqnarray}
c_1'&:=&l_0-\frac{h_0}{h_1^2}k_1^2+\sum_{i\not=1}\frac{1}{h_1^2|z_{i1}|}[2h_1k_1k_i-k_1^2h_i+h_1^2l_i],\\
 c_2'&:=&m_0+\frac{3}{2h_1^2}(h_1k_1l_0-h_0k_1l_1)-\frac{2h_0k_1^3}{h_1^3}\nonumber\\
&&+\sum_{i\not =1}\frac{1}{2h_1^3|z_{i1}|}[-(4k_1^3+3h_1k_1l_1)h_i+3h_1(2k_1^2+h_1l_1)k_i+3h_1^2k_1l_i+2h_1^3m_i].
\end{eqnarray}
The $1$-forms $\omega$ and $\chi$ behave, respectively,  as
\begin{eqnarray}
\hat\omega
&=&  \biggl[\sum_{j\not=1}\left(nm_j+\frac{3}{2}k_1 l_j- \frac{3}{2}k_jl_1\right)\frac{-\cos\theta}{|z_{j1}|} + \sum_{i,j\not=1(i\not=j)}\left(-m_j+\frac{3}{2}k_i l_j \right) \frac{z_{i1}z_{j1}}{|z_{i1}z_{j1}|z_{ji}}  \nonumber\\
&&-\left(m_0n+\frac{3}{2}l_0k_1-h_0m_1\right)\cos\theta -\sum_{i\not=1}\left(-m_0+\frac{3}{2}l_0k_i-h_0m_i\right) \frac{-z_{i1}}{|z_{i1}|}+c \biggr]d\phi,
\end{eqnarray}
and
\begin{eqnarray}
\chi=  h_1 \hat\omega_1+\sum_{i\not=1} h_i \hat\omega_i\simeq \left(n\cos\theta+\sum_{i\not=1}\frac{z_{i1}}{|z_{i1}|} \right)d\phi.
\end{eqnarray}
In terms of new coordinates $(v,\psi')$ defined by
\begin{eqnarray}
dv=dt-\left(\frac{A_0}{r^2}+\frac{A_1}{r}\right)dr,\qquad 
d\psi'=d\psi+\sum_{i\not=1}\frac{z_{i1}}{|z_{i1}|}d\phi-\frac{B_0}{r}dr,
\end{eqnarray}
we can confirm that the divergences of $g_{rr}$ and $g_{r\psi'}$ can be eliminated and the metric is analytic at ${\bm r}=0$, where the 
constants $A_0$, $A_1$, and $B_0$ are defined by
 \begin{eqnarray}
 {A_0}&=& \frac{1}{2} \sqrt{3k_1^2l_1^2+4h_1l_1^3-4m_1(2k_1^3+3h_1k_1l_1+h_1^2m_1)},\\
 {A_0B_0}&=& \frac{2  {k_1}^3+3 h_1 {k_1}  {l_1}+2h_1^2m_1 }{2}, \\
 {4A_0A_1}&=&3k_1^2l_0l_1+6h_1l_0l_1^2+2h_0l_1^3-4k_1^3m_0-4h_1^2m_0m_1-4h_0h_1m_1^2\notag\\
 &&-6k_1(h_1l_1m_0+h_1l_0m_1+h_0l_1m_1)\notag\\
 &+&\sum_{i\not=1}\frac{1}{|z_{i1}|}[(3k_1 l_1^2-12k_1^2m_1-6h_1l_1m_1)k_i+3(k_1^2 l_1 + 2h_1 l_1^2-2h_1k_1m_1 )l_i\notag\\
 &&-2(2 k_1^3 + 3h_1 k_1 l_1+2h_1^2m_1)m_i+2(l_1^3-3k_1l_1m_1-2h_1m_1^2)h_i]. 
  \end{eqnarray}
Hence, it turns out that the point ${\bm r}={\bm r}_1$ corresponds to the Killing horizon for the supersymmetric Killing field $V=\partial/\partial v$. 
Moreover,  after putting $(v,r)\to (v/\epsilon,\epsilon r)$,  taking the limit of $\epsilon \to 0$,  we can obtain the near-horizon geometry 
\begin{eqnarray}
ds^2_{\rm NH}&=&\frac{R_2^2}{4}\left[d\psi'+n\cos\theta d\phi -\frac{2k_1(2k_1^2+3nl_1)+4n^2m_1}{R_1^4R_2^2}rdv\right]^2+R_1^2(d\theta^2+\sin^2\theta d\phi^2)\nonumber \\
&&-\frac{4r^2}{R_1^2R_2^2}dv^2-\frac{4}{R_2}dvdr,
\label{NHmetric}
\end{eqnarray}
where we have defined 
\begin{eqnarray}
&&R_1^2:=k_1^2+nl_1,\label{sec:R1ineq}\\
&&R_2^2:=\frac{l_1^2(3k_1^2+4nl_1)-4m_1(2k_1^3+3nk_1l_1+n^2m_1)}{R_1^4}.\label{sec:R2ineq}
\end{eqnarray}
This metric is locally isometric to the near-horizon geometry of the Breckenridge-Myers-Peet-Vafa (BMPV) black hole~\cite{Chamseddine:1996pi}.
In order to eliminate CTCs around the horizon, we must require 
\begin{eqnarray}
R_1^2>0,\label{eq:R1^2-positive}
\end{eqnarray}
and
\begin{eqnarray}
R_2^2>0.\label{eq:R2^2-positive}
\end{eqnarray}
The metric of the spatial cross section of the event horizon can be read off as 
\begin{eqnarray}
ds^2_{\rm H}=\frac{R_2^2n^2}{4} \left(\frac{d \psi'}{n}+ \cos\theta d\phi\right)^2+R_1^2(d\theta^2+\sin^2\theta d\phi^2) \,, 
\end{eqnarray}
which is the squashed metric of the lens space $S^3/\mathbb Z_n$.




\subsection{The points ${\bm r}={\bm r}_i$ ($n=2,...,n$)}
 The metric of the Gibbons-Hawking space has apparent divergences at the points  ${\bm r}={\bm r}_i$ ($n=2,...,n$) but it can be  shown that they correspond to coordinate singularities under the appropriate parameter-setting. 
 We  impose the boundary conditions so that each point ${\bm r}={\bm r}_i$ ($n=2,...,n$)  behaves like the smooth origin of Minkowski spacetime.
Let us choose the coordinates $(x,y,z)$ on ${\mathbb E}^3$ in the Gibbons-Hawking space so that the $i$-th point ${\bm r}_i$ ($i\not=1$) is an origin of ${\mathbb E}^3$, 
near which ${\bm r}=0$, the functions $f^{-1}$ and $\omega_\psi$ behave, respectively, as
\begin{eqnarray}
f^{-1}&\simeq& \frac{l_i+\frac{k_i^2}{h_i}}{r}+c_1,\qquad 
\omega_\psi\simeq \frac{\frac{k_i^3}{h_i^2}+\frac{3}{2h_i}k_il_i+m_i}{r}+c_2,
\end{eqnarray}
where the constants $c_1$ and $c_2$ are defined by
\begin{eqnarray}
c_1&:=&l_0+\frac{-h_0k_i^2}{h_i^2}+\sum_{j(\not=i)}\frac{1}{h_i^2|z_{ji}|}(-k_i^2h_j+2h_ik_ik_j+h_i^2l_j), \\
c_2&:=&m_0+\frac{1}{2h_i^3}\left(-4h_0k_i^3+3h_i^2k_il_0-3h_0h_ik_il_i\right)\nonumber\\
&+&\sum_{j(\not =i)}\frac{1}{2h_i^3|z_{ji}|}[-(4k_i^3+3h_ik_il_i)h_j+3h_i(2k_i^2+h_il_i)k_j+3h_i^2k_il_j+2h_i^3m_j].\label{eq:c2}
\end{eqnarray}
The $1$-forms $\chi$ and $\hat\omega$ are approximated as
\begin{eqnarray}
\chi&\simeq& \left(h_i\cos\theta+\chi_{(0)}\right)d\phi\,, \qquad 
\hat\omega\simeq(\hat \omega_{(1)}\cos\theta+\hat\omega_{(0)})d\phi \,,\label{eq:1form}
\end{eqnarray}
where 
\begin{eqnarray}
\chi_{(0)} & := &- \sum_{j(\not=i)}\frac{h_jz_{ji}}{|z_{ji}|} \,,\\
\hat \omega_{(0)}& := &  \sum_{k,j(\not=i,k\not=j)}\left(h_km_j+\frac{3}{2}k_kl_j\right)\frac{z_{ji}z_{ki}}{|z_{ji}z_{ki}|z_{jk}}+\sum_{j(\not= i)}\left(m_0h_j+\frac{3}{2}k_jl_0-h_0m_j\right)\frac{z_{ji}}{|z_{ji}|}+c\,, \\
\hat \omega_{(1)}& := & -\sum_{j(\not=i)}\left(h_im_j-h_jm_i+\frac{3}{2}(k_il_j-k_jl_i)\right)\frac{1}{|z_{ji}|}-\left(m_0h_i+\frac{3}{2}k_il_0-h_0m_i\right)\,.
\end{eqnarray}
Therefore, the asymptotic behavior of the metric around this point can be written as
\begin{eqnarray}
ds^2&\simeq& -\left( \frac{k_i^2/h_i+l_i}{r}+c_1\right)^{-2}\biggl[dt+\left(\frac{k_i^3/h_i^2+\frac{3}{2}k_il/h_i+m_i/h_i^2}{r}+c_2\right)\left\{d\psi+(h_i\cos\theta+\chi_{(0)})d\phi \right\} \nonumber\\
&+&(\hat \omega_{(1)}\cos\theta+\hat\omega_{(0)})d\phi \biggr]^2+\left( \frac{k_i^2/h_i+l_i}{r}+c_1\right)\frac{r}{h_i}\biggl[\left\{d\psi+(h_i\cos\theta+\chi_{(0)})d\phi \right\}^2\nonumber\\
&+&h_i^2\left(\frac{dr^2}{r^2}+d\theta^2+\sin^2\theta d\phi^2\right)\biggr].
\end{eqnarray}

First, to remove the divergences in the functions $f^{-1}$ and $\omega_\psi$,  the following conditions  must be imposed on the parameters $(k_i,l_i,m_i)$
\begin{eqnarray}
l_i&=&-\frac{k_i^2}{h_i},\label{eq:condition1}\\
m_i&=&\frac{k_i^3}{2h_i^2}\label{eq:condition2},
\end{eqnarray}
which gives
\begin{eqnarray}
\frac{k_i^3}{h_i^2}+\frac{3k_il_i}{2h_i}+m_i=0,\qquad c_2=-h_i\hat\omega_{(1)}. \label{eq:condition12}
\end{eqnarray}
Introducing the new coordinates $(\rho,\psi', \phi')$ by
\begin{eqnarray}
\rho=2\sqrt{h_ic_1r}, \qquad  \psi'=\psi+\chi_{(0)}\phi, \qquad \phi'=\phi \,, 
\end{eqnarray}
we can obtain the metric near  ${\bm r}={\bm r}_i$, which is given by
\begin{eqnarray}
ds^2\simeq -c_1^{-2} d[t+c_2\psi'+\hat \omega_{(0)}\phi']^2+d\rho^2+\frac{\rho^2}{4}\left[(d\psi'+h_i\cos\theta d\phi')^2+d\theta^2+\sin^2\theta d\phi'{}^2\right] \,,
\label{nutmetric}
\end{eqnarray}
where to ensure that the metric has the Lorentzian signature, we have imposed
\begin{eqnarray}
h_ic_1>0, \quad i=2,...,n  \,. \label{eq:c1ineq}
\end{eqnarray}
Next, as explained in detail in Ref.~\cite{Tomizawa:2016kjh},  to remove the causal violation around each ${\bm r}_i$, we must impose that at ${\bm r}={\bm r}_i$ ($i=2,...,n$),
\begin{eqnarray}
&&c_2=0, \label{eq:c20}\\
&&\omega_{(0)}=0.
\end{eqnarray}
As shown below,  the so-called bubble equations~(\ref{eq:c20}) automatically guarantee $\hat\omega_{(0)}=0$ for all $i=2,...,n$, 
Therefore, each point ${\bm r}={\bm r}_i\ (i=2,\ldots,n)$ corresponds merely to the coordinate singularities like the origin of the Minkowski spacetime. 
Thus, we have shown that the points ${\bm r}={\bm r}_i\ (i=2,\ldots,n)$ describe the timelike and regular points. 

\medskip
Finally, we prove $\hat\omega_{(0)}=0$ holds at each ${\bm r}={\bm r}_i$  for $i=2,...,n$ . 
It can be shown from (\ref{eq:condition1}) and (\ref{eq:condition2}) that the bubble equations (\ref{eq:c20}) can be written as 
\begin{eqnarray}
0&=&m_0-\frac{3}{2}k_il_0+h_0m_i+\sum_{j(\not =i)}\frac{1}{|z_{ji}|}[3k_i^2k_j+2k_i^3h_j-\frac{3}{2}(k_il_j+l_ik_j+k_il_ih_j)+m_j]\notag \\
 &=&m_0-\frac{3}{2}k_i+h_0m_i-\frac{nk_i^3+3k_1k_i^2-3l_1k_i+2m_1}{2z_{1i}}+\sum_{2 \le j(\not =i)}\frac{(k_j-k_i)^3}{2|z_{ji}|}.\label{eq:c2=0}
 \end{eqnarray}
Furthermore,  the summation  of (\ref{eq:c2=0}) for $i=2,..., n$ gives
\begin{eqnarray}
 0&=&\sum_{2\le j}\left[m_0-\frac{3}{2}k_j+h_0m_j-\frac{nk_j^3+3k_1k_j^2-3l_1k_j+2m_1}{2z_{1j}}+\sum_{2 \le k(\not =i)}\frac{(k_k-k_j)^3}{2|z_{kj}|}\right] 
 \notag\\ 
&=&\sum_{2\le j}\left(m_0-\frac{3}{2}k_j+h_0m_j\right)-\sum_{2\le j }\frac{nk_j^3+3k_1k_j^2-3l_1k_j+2m_1}{2z_{1j}},\label{eq:sumc2=02}
\end{eqnarray}
where the last term in the first line vanishes by the antisymmetry for $k$ and $j$.
From Eqs.~(\ref{eq:condition1}) and (\ref{eq:condition2}), $\hat\omega_{(0)}$ is written as
{\small
\begin{eqnarray}
\hat\omega_{(0)}&=&\sum_{k,j(k,j\not=i,k\not=j)}\left(h_km_j+\frac{3}{2}k_kl_j\right)\frac{z_{ji}z_{ki}}{|z_{ji}z_{ki}|z_{jk}}+\sum_{j(\not=i)}\left( m_0h_j+\frac{3}{2}k_jl_0-h_0m_j\right)\frac{z_{ji}}{|z_{ji}|}\nonumber \\
&=&-\sum_{2\le j(\not=i)}\frac{nk_j^3-2h_j^3m_1-3h_jk_1k_j^2-3h_j^2k_jl_1}{2h_j^2z_{j1}}\frac{z_{ji}}{|z_{ji}|}+\sum_{2\le k,j (k,j \not=i,k\not=j)}\frac{h_kk_j^3-3h_jk_kk_j^2}{2h_j^2z_{jk}}\frac{z_{ji}z_{ki}}{|z_{ji}z_{ki}|}\nonumber\\
&&-\left(nm_0+\frac{3}{2}l_0k_1-h_0m_1\right)+\sum_{2\le j(\not=i)}\left( -m_0+\frac{3}{2}k_j-h_0m_j\right)\frac{z_{ji}}{|z_{ji}|}\notag \\
&&-\sum_{2\le j}\frac{nk_j^3-2h_j^3m_1-3h_jk_1k_j^2-3h_j^2l_1k_j}{2h_j^2z_{j1}}-\sum_{2\le k,j(k\not=j)}\frac{h_kk_j^3-3h_jk_kk_j^2}{2h_j^2z_{kj}}. \label{eq:omega}
\end{eqnarray}
}
The third, fifth and sixth terms of the right-hand side of (\ref{eq:omega}) are combined into
{\small
\begin{eqnarray}
&&-\left(nm_0+\frac{3}{2}k_1l_0-h_0m_1\right)-\sum_{2\le j}\frac{nk_j^3-2h_j^3m_1-3h_jk_1k_j^2-3h_j^2l_1k_j}{2h_j^2z_{j1}}-\sum_{2\le k,j(k\not=j)}\frac{h_kk_j^3-3h_jk_kk_j^2}{2h_j^2z_{kj}}\notag \\
&&~~=-\left(nm_0+\frac{3}{2}k_1l_0-h_0m_1\right)+\sum_{2\le j}\left(m_0-\frac{3}{2}k_jl_0+h_0m_j\right)+\sum_{2\le k,j (k\not=j)}\frac{(k_k-k_j)^3}{4z_{kj}}\notag \\
&&~~=\sum_{2\le j,k(k\not=j)}\frac{(k_k-k_j)^3}{4z_{kj}},\label{eq:356}
\end{eqnarray}
}
where we have used Eq.~(\ref{eq:sumc2=02}) for the second term in the first line and Eq.~(\ref{eq:m0}) for the last equality. 
Next, the summation of the first, second and fourth terms  on the right-hand
side of (\ref{eq:omega}) reduces to 
{\small
\begin{eqnarray}
&-&\sum_{2\le j(\not=i)}\frac{nk_j^3-2h_j^3m_1-3h_jk_1k_j^2-3h_j^2k_jl_1}{2h_j^2z_{j1}}\frac{z_{ji}}{|z_{ji}|}+\sum_{2\le k,j (k,j\not=i,k\not=j)}\frac{h_kk_j^3-3h_jk_kk_j^2}{2h_j^2z_{jk}}\frac{z_{ji}z_{ki}}{|z_{ji}z_{ki}|}\notag \\
&&+\sum_{2\le j(\not=i)}\left( -m_0+\frac{3}{2}k_j-h_0m_j\right)\frac{z_{ji}}{|z_{ji}|}\notag \\
&=&\sum_{2\le j\not=i}\frac{z_{ji}}{|z_{ji}|}\left[\left(-m_0+\frac{3}{2}k_j-h_0m_j+\frac{nk_j^3+3k_1k_j^2-3k_jl_1+2m_1}{2z_{1j}}\right)+\sum_{2\le k(\not=i,j)}\frac{(k_k-k_j)^3}{4z_{jk}}\frac{z_{ki}}{|z_{ki}|}\right]\notag \\
&=&\sum_{2\le j\not=i}\frac{z_{ji}}{|z_{ji}|}\left[\left(\sum_{2\le k(\not =j)}\frac{(k_k-k_j)^3}{2|z_{kj}|}\right)+\sum_{2\le k(\not=i,j)}\frac{(k_k-k_j)^3}{4z_{jk}}\frac{z_{ki}}{|z_{ki}|}\right]\notag \\
&=&\sum_{2\le j(\not=i)}\frac{z_{ji}}{|z_{ji}|}\left[\sum_{2\le k(\not=j)}\frac{(k_k-k_j)^3}{2|z_{kj}|}+\sum_{2\le k(\not=i,j)}\frac{(k_k-k_j)^3}{4z_{jk}}\frac{z_{ki}}{|z_{ki}|}\right],\\
&=&-\sum_{2\le j,k(k\not=j)}\frac{(k_k-k_j)^3}{4z_{kj}},
\label{eq:124}
\end{eqnarray}
}where we have used Eq.~(\ref{eq:sumc2=02}) for the second equality. Thus, the straightforward computations enables us 
to show that (\ref{eq:124}) coincides with (\ref{eq:356}) up to the minus sign. 
This completes the proof of $\hat \omega_{(0)}=0$.

\subsection{Axis}

The $z$-axis of ${\mathbb E}^3$ (i.e., $x=y=0$) in the Gibbons-Hawking space is split into  the $(n+1)$ intervals as $I_-=\{(x,y,z)|x=y=0,  z<z_1\}$, $I_i=\{(x,y,z)|x=y=0,z_i<z<z_{i+1}\}\ (i=1,...,n-1)$ and $I_+=\{(x,y,z)|x=y=0,z>z_n\}$. 
We find that on $I_\pm$, $\hat\omega$ vanishes since
\begin{eqnarray}
\hat\omega&=&\sum_{k,j(k\not=j)}\left(h_km_j+\frac{3}{2}k_kl_j\right)\hat\omega_{kj}-\sum_{j}\left(m_0h_j+\frac{3}{2}l_0k_j-h_0m_j\right)\hat\omega_j+c d\phi\notag\\
&=&\sum_{k,j(k\not=j)}\left(h_km_j+\frac{3}{2}k_kl_j\right)\frac{d\phi}{z_{jk}}\mp\sum_{j}\left(m_0h_j+\frac{3}{2}l_0k_j-h_0m_j\right)d\phi -\sum_{k,j(k\not=j)}\left(h_km_j+\frac{3}{2}k_kl_j\right)\frac{d\phi}{z_{jk}} \notag\\
&=&\mp\sum_{j}\left(m_0h_j+\frac{3}{2}l_0k_j-h_0m_j \right)d\phi \notag\\
&=&0,
\end{eqnarray}
where we have used Eq.~(\ref{eq:c}) and Eq.~(\ref{eq:m0}), respectively, in the second equality and the last equality.

For $z\in I_i\ (i=1,2,...,n-1)$, 
we find
\begin{eqnarray}
\hat \omega_{\phi}|_{I_i}&=&\hat \omega_{\phi}(\rho=0,z_i<z<z_{i+1})\\
                                        &=&\hat \omega_{\phi}|_{{\bm r}={\bm r}_i}(\theta=0)\\
                                        &=&\hat \omega_{(1)}+\hat \omega_{(0)}\\
                                        &=&-c_2\\
                                        &=&0
\end{eqnarray}
where we have used the fact that $\hat\omega$ is constant on $I_i$ in the second equality, and  Eq.~(\ref{eq:1form}) in the third equality. 
Furthermore, we have used Eq.~(\ref{eq:condition12}) and Eq.~(\ref{eq:c20}), respectively, in the fourth equality and last equality. 
It hence turns out that  $\hat\omega=0$ holds at each interval, which proves that no Dirac-Misner string pathologies exist throughout the spacetime. 
\medskip

In turns, to prove the absence of orbifold singularities, let us consider the rod structure.
On the interval $I_\pm$, we have
\begin{eqnarray}
\chi&=&\pm d\phi,
\end{eqnarray} 
and on the each interval $I_i$,  
\begin{eqnarray}
\chi
      &=&\left(n\frac{z-z_1}{|z-z_1|}-\sum_{2\le j\le i}\frac{z-z_j}{|z-z_j|}-\sum_{i+1\le j\le n-1}\frac{z-z_j}{|z-z_j|}\right)d\phi \notag \\
      &=&\left(2n-2i+1\right)d\phi.
\end{eqnarray}
Therefore, the two-dimensional $(\phi,\psi)$-part of the metric on the intervals $I_\pm$ and $I_i$
takes in the following simple form
\begin{eqnarray}
ds^2_2=(-f^2\omega_\psi+f^{-1}H^{-1})(d\psi+\chi_\phi d\phi)^2. \label{eq:axis}
\end{eqnarray}

We had better work in the coordinate basis  $(\partial_{\phi_1},\partial_{\phi_2})$ of  the periodicity of $2\pi$, 
rather than in $(\partial_\phi,\partial_\psi)$, where $(\phi_1,\phi_2)$ are defined by $\phi_1:=(\psi+\phi)/2$ and $\phi_2:=(\psi-\phi)/2$.  
From~(\ref{eq:axis}), we see that the rod vector is given by $v:=\partial_\phi-\chi_\phi\partial_\psi$ on each interval, which 
are explicitly written as
\begin{enumerate}
\item on $I_+$,  $v_+:=\partial_\phi-\partial_\psi=(0,-1)$
\item on each $I_i$ ($i=1,...,n-1$), $v_i:=\partial_\phi-(2n-2i+1)\partial_\psi=(i-n,i-n-1)$,
\item on $I_-$,  $v_-:=\partial_\phi+\partial_\psi=(1,0)$. 
\end{enumerate}
From these, we can observe that the rod vectors $v_\pm,\ v_i$ satisfy 
\begin{eqnarray}
{\rm det}\ (v_+^T,v_{n-1}^T)=-1,\qquad {\rm det}\ (v_{i}^T,v_{i-1}^T)=-1, 
\label{noorbifold}
\end{eqnarray}
with 
\begin{eqnarray}
{\rm det}\ (v_1^T,v_{-}^T)=n.
\label{lens_cond}
\end{eqnarray}
As mathematically shown in~\cite{Hollands:2007aj}, Eq.(\ref{noorbifold}) shows that  there exist no orbifold singularities at adjacent intervals $z=z_i\ (2\le i\le n)$, and Eq. (\ref{lens_cond}) shows that the horizon $z=z_1$ has the spatial topology of the lens space $L(n,1)=S^3/{\mathbb Z}_n$.




\section{Physical properties}
\label{sec:analysis}
Since appropriate boundary conditions are given in the last section, we can now investigate several physical properties of the solution obtained in Sec.~\ref{sec:solution}. 
To to do,  we can consider the physical conserved charges from two points of view, in the five-dimensional minimal supergravity and in the dimensionally reduced four-dimensional theory, which leads to a massless axion and a dilaton coupled to gravity and two $U(1)$ gauge field, one of which has Chern-Simon coupling.  Here, let us take the five-dimensional point of view for the simplicity. 
 Since at infinity $r\to\infty$ the spacetime asymptotically behaves as an $S^1$ fiber bundle over four-dimensional Minkowski spacetime, whose metric can be written as $g_{\mu\nu}\simeq\eta_{\mu\nu}+h_{\mu\nu}$ in Cartesian coordinates, 
 the ADM mass and ADM (angular) momentum can be computed.  
  Following the notations in Ref.~\cite{Elvang:2005sa}, we can express the $2\times2$ ADM stress tensor as
\begin{eqnarray}
T_{ab}=\frac{1}{16\pi G_5}\int_{S_\infty}d\Omega_{S^2}^2r^2n^i[\eta_{ab}(\partial_ih^{c}{}_c-\partial_jh^j{}_i)-\partial_ih_{ab}], \quad(a,b,c =\bar t,\bar w, \ i,j=x,y,z)
\end{eqnarray}
where $d\Omega_{S^2}^2$ is a volume element of a two-dimensional sphere with unit radius and $n^i$ is the radial unit normal vector.   
In terms of the stress tensor, we have the ADM mass and ADM (angular) momentum  along the fifth dimension $\partial/\partial\bar w$, respectively, as
{\small
\begin{eqnarray}
M&=&\int d\bar wT_{\bar t\bar t}\nonumber\\
&=&\frac{\pi R_k}{2G_5}\frac{(1-2h_0m_0^2)\sum_ih_i+h_0(3\sum_i l_i-3m_0\sum_i k_i-2h_0m_0 \sum_i m_i)}{1-h_0m_0^2},\nonumber\\
P&=&\int d\bar wT_{\bar t\bar w}\nonumber\\
&=&\frac{\pi R_k}{4G_5}\frac{2m_0\sum_i h_i-6h_0m_0\sum_i l_i+(1+h_0m_0^2)(3\sum_ik_i+2h_0\sum m_i)}{1-h_0m_0^2}\nonumber.
\end{eqnarray}
}
Moreover, the angular momentum along $\partial/\partial\phi$ can be obtained as
\begin{eqnarray}
J_\phi
=\frac{\pi R_5}{G_5}\sum_i\left(m_0h_i+\frac{3}{2}k_i-h_0m_0\right)z_i.
\end{eqnarray}
As pointed out in Refs.~\cite{Kunduri:2014kja,Tomizawa:2016kjh}, let us note that the asymptotically flat supersymmetric black lens must have two non-zero angular momenta. 
Now we would like to see whether the Kaluza-Klein black lens obtained here allows two zero-angular momenta or not, in particular, for the simplest case of $n=2$.  
From Eq.~(\ref{eq:c20}), $z_{21}$ can be written in terms of the other parameters, as
 \begin{eqnarray}
z_{21}=\frac{2k_2^3+3k_1k_2^2-3l_1k_2+2m_1}{-2h_0k_2^3+6k_2+3k_1-2h_0m_1},
\end{eqnarray}
and the substitution of this into the inequality~(\ref{eq:c1ineq}) yields
\begin{eqnarray}
(c_1=)\ 1-h_0k_2^2+\frac{({-2h_0k_2^3+6k_2+3k_1-2h_0m_1})(l_1-2k_2^2-2k_1k_2)}{2k_2^3+3k_1k_2^2-3l_1k_2+2m_1}<0. \label{eq:c1ineq2}
\end{eqnarray}
The solid curve in FIG.1 shows the plots of $P=0$ in the $(k_1,k_2)$-plane for $n=2$, $z_1=0$, $h_0=m_1=l_1=1$. 
The shaded portions in this figure present the region such that all of the inequalities $R_1^2>0$, $R_2^2>0$, $z_{21}>0$ and $c_1<0$ are satisfied. 
It can be seen from this figure that there indeed exists a parameter region in which $P=0$ can be realized. 
Moreover, it can be shown from~Eq.(\ref{eq:m0}) that  for $n=2$ the angular momentum $J_\phi$ vanishes in the choice of the parameter $z_1=0$. 
Therefore, at least, for $n=2$, in contrast to the result of the asymptotically flat supersymmetric black lens, we can see that there exists a case where both of angular momenta $(J_\phi,P)$ vanish.  

\begin{figure}[t]
\label{fig:P0}
\begin{center}
\includegraphics[width=8cm]{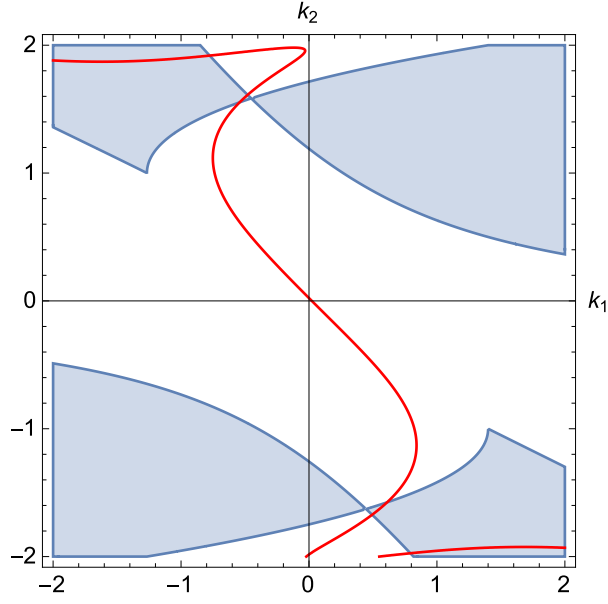}
\caption{Plots of $P=0$ in a $(k_1,k_2)$-plane for $n=2$, $z_1=0$, $h_0=m_1=l_1=1$. 
The shaded regions and the curve denote the region of $R_1^2>0$, $R_2^2>0$, $c_1<0$, $z_{21}>0$ and $P=0$, respectively. 
The  (angular) momentum $P$ vanishes where the curve crosses on the shaded region.
}
\end{center}
\end{figure}

The  interval $I_1$ is topologically a disc and the $(n-2)$ intervals $I_i$ ($i=2,...,n-1$) is a two-dimensional sphere. 
The magnetic fluxes through $I_i$ which are defined by 
\begin{eqnarray}
q[I_i]:=\frac{1}{4\pi}\int_{I_i}F\,
\end{eqnarray}
are computed as
\begin{eqnarray}
q[I_1]=\frac{\sqrt{3}}{2}\left[
\frac{k_1l_1}{2(k_1^2+nl_1)}
-k_2
\right]\,, \qquad 
q[I_i]=\frac{\sqrt{3}}{2}(k_i-k_{i+1})~~ (i=2,... n-1). 
\label{mag_fluxes}
\end{eqnarray}
In particular, the first term in $q[I_1]$ gives the contribution from the horizon, whereas the second term and each term in $q[I_i]\ (i=2,\ldots,n)$ come from ${\bm r}={\bm r}_i\ (i=2,\ldots,n)$. 
The expression~(\ref{mag_fluxes}) for the magnetic fluxes are exactly the same as for the asymptotically flat black lens with the horizon topology of $L(n,1)$ in Ref.~\cite{Tomizawa:2016kjh}.
As shown in Refs.~\cite{Kunduri:2014kja,Tomizawa:2016kjh}, for the asymptotically flat supersymmetric black lenses, the existence of the magnetic fluxes plays an essential role in supporting the horizon of the lens space topology.  
On the contrary, it can be shown that this is not true for the Kaluza-Klein supersymmetric black lens obtained here.
In turns, we consider whether the Kaluza-Klein black lens also prohibits $(q[I_1],\ldots,q[I_{n-1}])= (0,\ldots,0)$.
For $k_2=k_1l_1/[2(k_1^2+nl_1)]$, $k_i=k_{i+1}\ (i=2,...,n-1)$, all magnetic fluxes $q[I_i]\ (i=1,...,n-1)$ vanish. 
In the choice of these parameters, the condition~(\ref{eq:c1ineq}) can be simply written as 
\begin{eqnarray}
-\frac{nl_1^2(3k_1^2+4nl_1)}{4(k_1^2+nl_1)^2}>(1-h_0k_i^2)z_{i1}, \qquad i=2,...,n.
\end{eqnarray}
As exactly proved in Ref.~\cite{Tomizawa:2016kjh} for the asymptotically flat black lens, which can be realized by putting $h_0=m_1=0$, these inequalities cannot be satisfied, since the left-hand side is non-positive by Eq.~(\ref{eq:R2^2-positive}) but the right-hand side must be positive from our assumption. 
However, for the Kaluza-Klein black lens with $h_0>0$, the right-hand side can be negative due to the existence of the constant $h_0$, which corresponds to the size of an extra-dimension at infinity. 
In fact, we can see from FIG.2 that the magnetic flux can vanish, at least, for $n=2$.
The solid curve in FIG.2 denotes the plots of $q[I_1]=0$ in the $(k_1,k_2)$-plane for $n=2$, $z_1=0$, $h_0=10^{5}, m_1=10^{-4}, l_1=1$, and the two separated shaded portions are the regions such that all of the inequalities $R_1^2>0$, $R_2^2>0$, $z_{21}>0$ and $c_1<0$ are satisfied. 
Therefore, the magnetic flux $q[I_1]$ vanishes on the solid curve in the shaded regions.
Thus, in contrast to the asymptotically flat supersymmetric black lens, the magnetic flux can vanish for the Kaluza-Klein supersymmetric black lens.

\begin{figure}[t]
\label{fig:q10}
\begin{center}
\includegraphics[width=8cm]{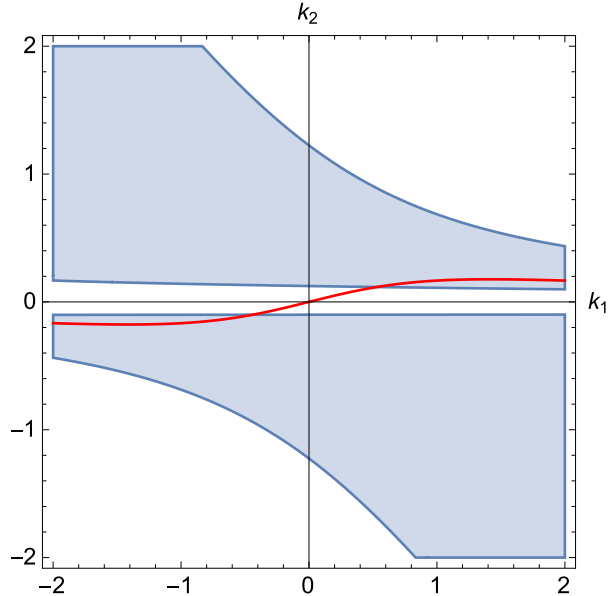}
\caption{
Plots of $q[I_1]=0$ in the $(k_1,k_2)$-plane for $n=2$, $z_1=0$, $h_0=10^{5}, m_1=10^{-4}, l_1=1$. The shaded regions and the curve denote the region of $R_1^2>0$, $R_2^2>0$, $z_{21}>0$ $c_1<0$ and $q[I_1]=0$, respectively. The magnetic flux $q[I_1]$ vanishes on the curve in the shaded regions.
}
\end{center}
\end{figure}




\section{Summary}
\label{sec:discuss}
In this work, we have constructed a  bi-axisymmetric Kaluza-Klein black lens solution as a supersymmetric solution in the bosonic sector of the five-dimensional minimal supergravity. 
We have shown that the spacetime has a degenerate Killing horizon with the spatial cross section of the lens topology of $L(n,1)=S^3/{\mathbb Z}_n$, and  also computed the mass, two angular momenta, and $(n-1)$ magnetic fluxes. 
When the compactification radius of the extra dimension becomes infinite, this solution exactly coincides with the asymptotically flat black lens in the previous work~\cite{Kunduri:2014kja,Tomizawa:2016kjh}.

\medskip
For the asymptotically flat supersymmetric black lens in Refs.~\cite{Kunduri:2014kja,Tomizawa:2016kjh} which can be obtained by taking the limit $h_0\to0$,  a pair of angular momenta cannot vanish, 
whereas for the Kaluza-Klein black lens in this paper,  both of them can vanish at least for $n=2$.
 For the asymptotically flat black lens, the existence of the magnetic fluxes plays an essential role in supporting the horizon of the black lens, whereas for the Kaluza-Klein black lens obtained in this paper,  this cannot be applied since the magnetic flux vanishes, at least, for $n=2$.




\acknowledgments
This work was supported by the Grant-in-Aid for Scientific Research (C) (Grant Number ~17K05452) from the Japan Society for the Promotion of Science.




\end{document}